\documentclass[prb,twocolumn,showpacs]{revtex4}
\usepackage{graphicx} 

\usepackage{amsmath,amsthm,amssymb,mathrsfs}
\usepackage{bm}

\begin{document}

\title{Theoretical condition for switching the magnetization in a perpendicularly magnetized ferromagnet via the spin Hall effect}

\author{Tomohiro Taniguchi
      }
 \affiliation{
 ${}^{1}$National Institute of Advanced Industrial Science and Technology (AIST), Spintronics Research Center, Tsukuba, Ibaraki 305-8568, Japan
 }

\date{\today} 
\begin{abstract}
{
A theoretical formula is derived for the threshold current to switch a perpendicular magnetization in a ferromagnet by the spin Hall effect. 
The numerical simulation of the Landau-Lifshitz-Gilbert equation indicates that magnetization switching is achieved when the steady-state solution of the magnetization 
in the presence of the current is outside an energetically unstable region. 
Based on the numerical result, an analytical theory deriving the threshold current is developed by focusing on the first-order perturbation to the unstable state. 
The analytical formula clarifies that the magnitude of the magnetic field applied to the current direction should be larger than 15\% of the perpendicular magnetic anisotropy field, 
and the current is less than the derived threshold value. 
}
\end{abstract}

 \maketitle



\section{Introduction}
\label{sec:Introduction}

The theoretical predictions of the spin-transfer torque effect in nanostructured ferromagnetic/nonmagnetic multilayers 
have drastically changed our understanding of the electron transport and magnetization dynamics in the ferromagnet [\onlinecite{slonczewski96,berger96}]. 
It provides various interesting physics in condensed matter physics and nonlinear science, 
such as spin-dependent transport in fine structures [\onlinecite{hillebrands06}] and a switching and limit cycle of magnetization [\onlinecite{bertotti09text}]. 
In addition, since the current necessary to excite the dynamics decreases with decreasing the volume of the ferromagnet, 
the spin-transfer torque effect has also attracted much attention from an applied physics viewpoint [\onlinecite{dieny16}]. 
The magnetization switching by the spin-transfer torque is particularly important for both fundamental and applied physics. 
Whereas it had been first confirmed in current-perpendicular-to-plane metallic [\onlinecite{katine00}] and highly resistive [\onlinecite{kubota05}] systems, 
the magnetization switching in a current-in-plane system was also demonstrated recently [\onlinecite{liu12,pai12,cubukcu14,torrejon15,fukami16}], 
where the spin-transfer torque is excited by spin current generated by the spin Hall effect in nonmagnetic heavy metals [\onlinecite{dyakonov71,hirsch99,kato04}]. 


A key quantity for the magnetization switching is the threshold current. 
Its theoretical formula for various systems has been derived 
by solving the Landau-Lifshitz-Gilbert (LLG) equation using several approaches such as
linearization [\onlinecite{sun00,grollier03,morise05,taniguchi15}], the averaging technique [\onlinecite{hillebrands06,taniguchi15PRB,taniguchi16}], or integration [\onlinecite{yamada15,taniguchi19}]. 
In particular, the threshold current formula for the switching of a perpendicular magnetization by the spin Hall effect was derived in Ref. [\onlinecite{lee13}], 
where the formula was obtained by analyzing a steady-state solution of the magnetization. 
Simultaneously, however, it was clarified in Ref. [\onlinecite{lee13}] that the derived formula is limitedly applicable to the ferromagnet with a large damping constant $\alpha( \gtrsim 0.03)$ 
because of the complex dependence of the threshold current on the damping constant for a small $\alpha$. 
This fact indicates the necessity of solving two issues. 
The first one is to clarify the origin of the complex dependence of the threshold current on $\alpha$. 
The second one is an extension of the theory to a small $\alpha$ limit 
because typical ferromagnetic materials, such as CoFeB, used in the spintronics devices have small damping constant on the order of $10^{-3}$ [\onlinecite{oogane06,konoto13,tsunegi14}]. 


In this paper, both numerical simulation and analytical theory are performed on the switching of the perpendicularly magnetized ferromagnet by the spin Hall effect. 
The phase diagram of the magnetization state are obtained by the numerical simulation of the LLG equation. 
The result indicates that the complex dependence of the switching current appears when the steady state of the magnetization in the presence of the current is in an energetically unstable region. 
It is clarified that although the formula in Ref. [\onlinecite{lee13}] is still applicable to the small damping region, 
there is another boundary of the current to achieve the switching with high accuracy. 
Based on the numerical results, the theoretical formula of another threshold current is derived. 
The new formula indicates that the magnetization switching occurs when the magnetization field larger than 15\% of the perpendicular magnetic anisotropy field is applied to the longitudinal direction, 
combined with the fact that the magnitude of the current is in the range between the threshold currents derived in Ref. [\onlinecite{lee13}] and the present work. 





\begin{figure}
\centerline{\includegraphics[width=1.0\columnwidth]{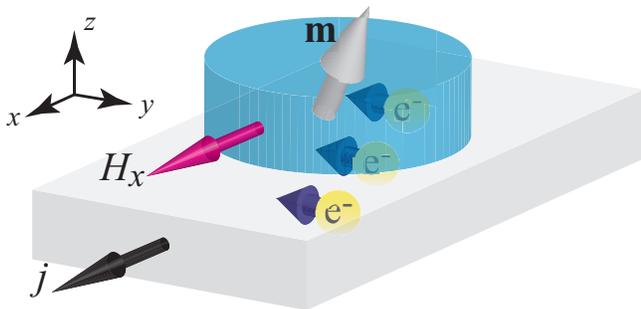}}
\caption{
         Schematic view of the system. 
         The ferromagnet having the magnetization vector $\mathbf{m}$ ($|\mathbf{m}|=1$) is placed on the nonmagnet. 
         The current density $j$ and external field $H_{x}$ are applied to the bottom nonmagnet and the ferromagnet, respectively, in the $x$ direction. 
         The spin current having the spin polarization along the $y$ direction is injected into the ferromagnet by the spin Hall effect. 
         \vspace{-3ex}}
\label{fig:fig1}
\end{figure}




\section{System description and numerical results}
\label{sec:System description and numerical results}

In this section, we describe the system under consideration, 
review the previous work briefly, and show the results of the numerical simulations. 


\subsection{System description}
\label{sec:System description}

The system we consider is schematically shown in Fig. \ref{fig:fig1}.
The ferromagnetic layer having the magnetization $\mathbf{m}$ ($|\mathbf{m}|=1$) is placed on the nonmagnetic heavy metal. 
The $z$ axis is perpendicular to the plane, whereas the $x$ axis is parallel to the direction of the electric current density $j$ in the nonmagnet. 
An external field $H_{x}$ is also applied to the ferromagnet along the longitudinal ($x$) direction. 
The magnetization dynamics in the ferromagnet is described by the LLG equation, 
\begin{equation}
  \frac{d \mathbf{m}}{dt}
  =
  -\gamma
  \mathbf{m}
  \times
  \mathbf{H}
  -
  \gamma
  H_{\rm s}
  \mathbf{m}
  \times
  \left(
    \mathbf{e}_{y}
    \times
    \mathbf{m}
  \right)
  +
  \alpha
  \mathbf{m}
  \times
  \frac{d \mathbf{m}}{dt},
  \label{eq:LLG}
\end{equation}
where $\gamma$ and $\alpha$ are the gyromagnetic ratio and the Gilbert damping constant, respectively. 
The magnetic field is given by 
\begin{equation}
  \mathbf{H}
  =
  H_{x}
  \mathbf{e}_{x}
  +
  H_{\rm K}
  m_{z}
  \mathbf{e}_{z}, 
  \label{eq:field}
\end{equation}
where $H_{\rm K}$ is the net magnetic anisotropy field along the $z$ direction. 
Since we are interested in a perpendicularly magnetized ferromagnet, $H_{\rm K}$ should be positive. 
For the latter discussion, it is useful to introduce the magnetic energy density $E$ as 
$E=-M \int d \mathbf{m}\cdot\mathbf{H}$, where $M$ is the saturation magnetization. 
Note that the energy density $E$ has two minima corresponding to $\mathbf{m}_{0+}=(\sin\Theta,0,\cos\Theta)$ 
and $\mathbf{m}_{0-}=(\sin\Theta,0,-\cos\Theta)$, where $\Theta=\sin^{-1}(|H_{x}|/H_{\rm K})$. 
Throughout this paper, we assume that the initial state of the magnetization is located at $\mathbf{m}_{0+}$, for convention, 
which is close to $+\mathbf{e}_{z}$, whereas $\mathbf{m}_{0-}$ is close to $-\mathbf{e}_{z}$. 
The strength of the spin-transfer torque by the spin Hall effect is 
\begin{equation}
  H_{\rm s}
  =
  \frac{\hbar \vartheta j}{2eMd},
\end{equation}
where $\vartheta$ is the spin Hall angle of the nonmagnetic heavy metal, 
whereas $d$ is the thickness of the free layer. 
The values of the parameters used in this work are derived from Ref. [\onlinecite{taniguchi15}] as 
$M=1500$ emu/c.c., $H_{\rm K}=540$ Oe, $\vartheta=0.1$, $d=1.0$ nm, $\gamma=1.764 \times 10^{7}$ rad/(Oe s), and $\alpha=0.005$. 
Note that the value of the damping constant is similar to that of CoFeB [\onlinecite{konoto13,tsunegi14}], 
and is one order of magnitude smaller than the value assumed in the previous theoretical analysis [\onlinecite{lee13}]. 


\subsection{Brief review of previous work}
\label{sec:Brief review of previous work}

For the following discussion, it is useful to briefly review the derivation of the threshold current in Ref. [\onlinecite{lee13}]. 
Reference [\onlinecite{lee13}] uses the fact that the solution of the LLG equation in the presence of the current finally saturates to a fixed point. 
In terms of the zenith and azimuth angles $(\theta,\varphi)$, defined as $\mathbf{m}=(\sin\theta\cos\varphi,\sin\theta\sin\varphi,\cos\theta)$, 
the LLG equation in a steady state is given by 
\begin{equation}
  H_{x}
  \sin\varphi
  +
  H_{\rm s}
  \cos\theta
  \sin\varphi
  =
  0,
  \label{eq:condition_1}
\end{equation}
\begin{equation}
  H_{\rm K}
  \sin\theta
  \cos\theta
  -
  H_{x}
  \cos\theta
  \cos\varphi
  -
  H_{\rm s}
  \cos\varphi
  =
  0.
  \label{eq:condition_2}
\end{equation}
It is also assumed in Ref. [\onlinecite{lee13}] that the magnetization moves from the stable state to the $x$ direction where $m_{y}=0$, or equivalently, $\varphi=0$,  
when the current is injected. 
Then, Eq. (\ref{eq:condition_1}) becomes self-evident, whereas Eq. (\ref{eq:condition_2}) gives a threshold current 
\begin{equation}
\begin{split}
  j_{\rm c\pm}
  =
  \pm
  &
  \frac{2eMd}{\hbar \vartheta}
  H_{\rm K}
  \frac{\mp 3 h + \sqrt{8+h^{2}}}{16}
\\
  & \times
  \sqrt{
    8
    \mp
    2 h 
    \left(
      \pm h 
      +
      \sqrt{
        8
        +
        h^{2}
      }
    \right)
  },
  \label{eq:jc}
\end{split}
\end{equation}
where $h=H_{x}/H_{\rm K}$. 
The physical meaning of the threshold here is that the magnetization cannot stay in a hemisphere 
with $m_{y}=0$ when the current magnitude exceeds Eq. (\ref{eq:jc}); see Appendix \ref{sec:AppendixA}.



\begin{figure*}
\centerline{\includegraphics[width=2.0\columnwidth]{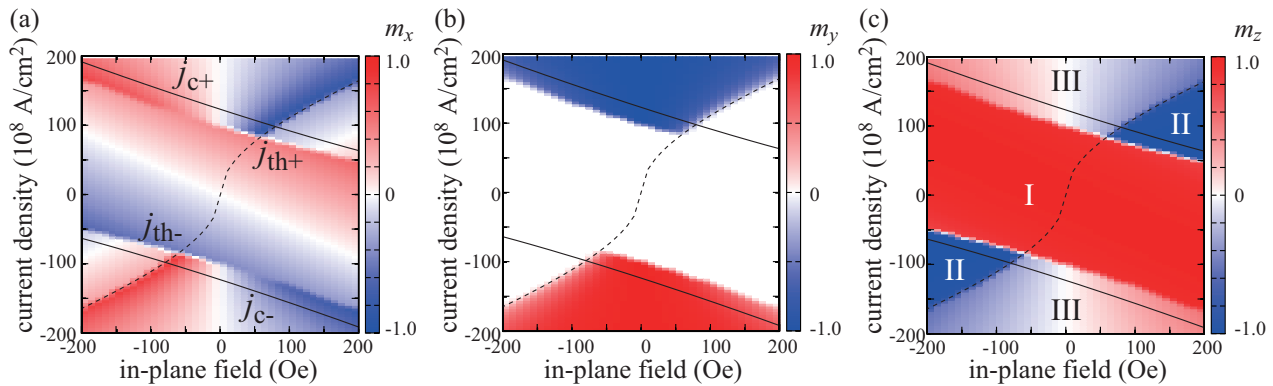}}
\caption{
        The steady state solutions of (a) $m_{x}$, (b) $m_{y}$, and (c) $m_{z}$ as functions of the in-plane magnetic field $H_{x}$ and the current density $j$. 
        The dynamical phase is divided into three regions, labeled as I, II, and III in (c). 
        The theoretical formulas of the threshold current, Eq. (\ref{eq:jc}) and (\ref{eq:jth}), are shown by the solid and dotted lines; see labels in (a). 
         \vspace{-3ex}}
\label{fig:fig2}
\end{figure*}



\subsection{Numerical simulation}
\label{sec:Numerical simulation}

Now let us show the numerical solutions of Eq. (\ref{eq:LLG}). 
Figures \ref{fig:fig2}(a), \ref{fig:fig2}(b), and \ref{fig:fig2}(c) are the phase diagram of 
$m_{x}$, $m_{y}$, and $m_{z}$, respectively. 
These steady-state solutions, satisfying $d \mathbf{m}/dt=\bm{0}$, are determined by 
the first and second terms on the right-hand side of Eq. (\ref{eq:LLG}). 
When the current magnitude is small and thus, the magnetization stays near the initial state, 
the magnetization lies in the $xz$ plane 
because $\mathbf{H}+H_{\rm s}\mathbf{e}_{y}\times\mathbf{m}$ in Eq. (\ref{eq:LLG}) becomes zero in case the magnetization is in the plane. 
On the other hand, when the current magnitude is large and thus the magnetization moves far away from the $z$ axis, 
$\mathbf{m}$ moves to the $y$ direction because the spin polarization of the incoming spin current points to this direction. 


The phase diagram in Fig. \ref{fig:fig2} is divided into three regions, labeled I, II, and III in Fig. \ref{fig:fig2}(c). 
The first one locates near the zero-current region, where the magnetization stays close to the initial state, $m_{z}\simeq 1$. 
The second region appears when the applied field exceeds a certain value, which is about 80 Oe, 
in which the magnetization moves close to the switched state, $m_{z} \simeq -1$. 
The third region corresponds to the other region where the magnetization stays near the $xy$ plane, where $m_{z} \simeq 0$. 
The solid lines in Fig. \ref{fig:fig2} correspond to those of Eq. (\ref{eq:jc}). 
It can be seen from these figures that the formula, Eq. (\ref{eq:jc}), provides a reasonable estimation 
of the lower boundary of the instability threshold, i.e., 
the magnetization stays in the first region, which is near the initial stable state ($m_{z}\simeq 1$), when the current magnitude is less than Eq. (\ref{eq:jc}). 
However, the formula cannot distinguish between the second ($m_{z}\simeq -1$) and third ($m_{z} \simeq 0$) regions. 



\begin{figure}
\centerline{\includegraphics[width=1.0\columnwidth]{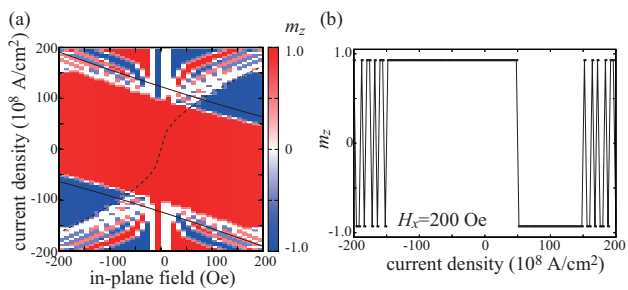}}
\caption{
        (a) The steady state of $m_{z}$ after turning off the current. 
        The field $H_{x}$ is fixed to $200$ Oe in (b). 
         \vspace{-3ex}}
\label{fig:fig3}
\end{figure}



We should note that the experimental study often investigates the magnetization state after turning off the current. 
Therefore, we also attempt to calculate the relaxed state of the magnetization after the current is turned off, as shown in Fig. \ref{fig:fig3}(a). 
It is revealed that the magnetization switches to the stable point $\mathbf{m}_{0-}$ 
when the magnetization in the presence of the current stays in region II in Fig. \ref{fig:fig2}(c). 
On the other hand, when the magnetization was in region III, 
the relaxed state after turning off the current becomes either $\mathbf{m}_{0+}$ and $\mathbf{m}_{0-}$, depending on the values of $j$ and $H_{x}$. 
Figure \ref{fig:fig3}(b) illustrates such deterministic and complex switching behavior by showing the magnetization state as an example, 
in the case where $H_{x}$ is 200 Oe in Fig. \ref{fig:fig3}(a). 
Here, for the current density of $-148 \lesssim j \lesssim 48$ MA/cm${}^{2}$, the magnetization returns to the initial state $\mathbf{m}_{0+}$. 
On the other hand, for the current density of $48 < j \lesssim 148$ MA/cm${}^{2}$, the magnetization definitely switches to the other stable state $\mathbf{m}_{0-}$. 
However, outside these regions, the magnetization relaxes to either $\mathbf{m}_{0+}$ and $\mathbf{m}_{0-}$. 
Such a complex behavior of the relaxed state will be an origin of a back-hopping in a high current region, 
which is recently studied in a current-perpendicular-to-plane system [\onlinecite{abert18,safranski19}]. 




\begin{figure*}
\centerline{\includegraphics[width=2.0\columnwidth]{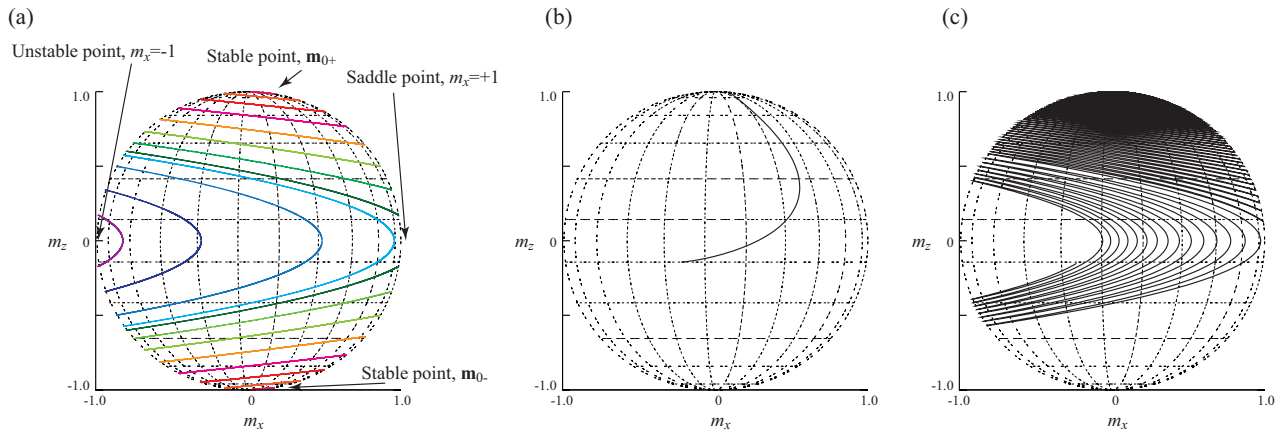}}
\caption{
         (a) Energy landscape of the perpendicular ferromagnet in the presence of the external field $H_{x}$ of $50$ Oe along the $x$ direction.
         (b) The dynamic trajectory of the magnetization in the presence of the current density $j=160$ MA/cm${}^{2}$. 
             Starting from the stable state $\mathbf{m}_{0+}$, the magnetization arrives at a point in an energetically unstable state. 
         (c) The relaxation dynamics of the magnetization after the current is turned off. 
             Staring from the steady state in (b), the magnetization finally returns to the stable state, $\mathbf{m}_{0+}$. 
         \vspace{-3ex}}
\label{fig:fig4}
\end{figure*}



\subsection{Switching mechanism}
\label{sec:Switching mechanism}

To apprehend such a complex dependence of the relaxation state on the current and field, 
it is useful to study the dynamics on the energy landscape of the energy density $E$. 
Figure \ref{fig:fig4}(a) shows the energy landscape of the present system, where the lines correspond to the constant energy curves of $E$. 
There are two stable states, $\mathbf{m}_{0+}$ and $\mathbf{m}_{0-}$, near $\pm \mathbf{e}_{z}$. 
The highest energy state is located at $\mathbf{m}_{\rm u}\equiv -\mathbf{e}_{x}$. 
On the other hand, the point $\mathbf{m}_{\rm s}\equiv +\mathbf{e}_{x}$ corresponds to the saddle point. 
For the sake of convention, let us call the regions between $\mathbf{m}_{0\pm}$ and $\mathbf{m}_{\rm s}$ the stable regions 
and the region between $\mathbf{m}_{\rm s}$ and $\mathbf{m}_{\rm u}$ the unstable region. 
Note that there are two boundaries between the stable and unstable regions because of the presence of two stable fixed points, $\mathbf{m}_{0\pm}$. 
We emphasize that the complex switching behavior mentioned above appears when the steady-state solution in the presence of the current is located in the unstable region. 
Figures \ref{fig:fig4}(b) and \ref{fig:fig4}(c) show an example of such dynamics. 
In Fig. \ref{fig:fig4}(b), the dynamic trajectory of the magnetization in the presence of a current density of $j=160$ MA/cm${}^{2}$ and longitudinal field $H_{x}=50$ Oe is shown. 
The magnetization finally locates at a certain point inside the unstable region. 
After the current is turned off, the magnetization starts to precess around the negative $x$ direction, as shown in Fig. \ref{fig:fig4}(c). 
This is because the torque due to the magnetic field induces the precession on a constant energy curve. 
Because of the presence of the damping torque, however, the magnetization slowly relaxes to the lower energy state. 
In the case of Fig. \ref{fig:fig4}(c), the magnetization traverses the upper ($m_{z}>0$) boundary and therefore is saturated to $\mathbf{m}_{0+}$. 
In this case, the switching does not occur. 


The result shown in Fig. \ref{fig:fig4} reveals the reason why the large damping assumption was necessary in the previous work [\onlinecite{lee13}]. 
Let us consider the case that the steady-state solution in the presence of the current locates inside the unstable region, as shown in Fig. \ref{fig:fig4}(b). 
When the damping constant is small, as in the case in Fig. \ref{fig:fig4}(c), the magnetization undergoes many precessional oscillation 
before traversing the boundary between the stable and unstable regions. 
Which of the relaxed states, $\mathbf{m}_{0-}$ or $\mathbf{m}_{0+}$, is accomplished is determined by 
whether the magnetization traverses the lower or upper boundary between the stable and unstable region. 
It depends on many parameters in the LLG equation, such as the damping constant and the longitudinal field, 
as well as the steady state solution in the presence of the current, which is the initial condition of the relaxation dynamics. 
Therefore, the current and/or field dependence of the relaxation dynamics becomes complex, as can be observed in Fig. \ref{fig:fig3}. 
Although the relaxed state can be predicted deterministically from the LLG equation in principle, 
it is difficult to obtain the analytical solution due to the nonlinearity of the LLG equation. 
On the other hand, when the damping constant is large, the relaxation dynamics becomes fast. 
In this case, the magnetization will traverse the boundary between the stable and unstable state 
without showing the precession around the negative $x$ axis. 
Then, the magnetization switches its direction accurately. 


At the end of this section, we should mention that the complex behavior of the relaxed state is not related to chaos, contrary to the suggestion in Ref. [\onlinecite{lee13}]. 
As can be seen in the derivation of Eq. (\ref{eq:jc}) above, 
the magnetization dynamics in the present system is described by two dynamical variables, $\theta$ and $\varphi$. 
On the other hand, chaos is prohibited in a two-dimensional system, according to the Poincar\'e-Bendixson theorem [\onlinecite{strogatz01}]. 
Therefore, the complex dependence of the switched state on the current and/or field cannot be explained by chaos theory. 


\section{Theoretical conditions for switching}
\label{sec:Theoretical conditions for switching}

The above discussion indicates the existence of another threshold current density to achieve the switching. 
As mentioned, the complex switching behavior appears when the steady state solution of the magnetization 
in the presence of the current locates inside the energetically unstable region. 
In fact, comparing Figs. \ref{fig:fig2}(b), \ref{fig:fig2}(c), and \ref{fig:fig3}(a), 
we notice that the complex switching behavior appears in region III, where $m_{y} \neq 0$, corresponding to the unstable region. 
On the other hand, magnetization switching occurs when the steady state solution satisfies $m_{y} \simeq 0$, as shown in region II in Fig. \ref{fig:fig2}. 
Therefore, we will focus on a small perturbation, $|\delta\varphi| \ll 1$, in Eqs. (\ref{eq:condition_1}) and (\ref{eq:condition_2}), 
where $\delta\varphi$ is a small deviation from $\varphi=0,\pi$. 
Then, we obtain 
\begin{equation}
  \left(
    H_{x}
    +
    H_{\rm s}
    \cos\theta
  \right)
  \delta
  \varphi
  =
  0,
  \label{eq:condition_1_rev}
\end{equation}
\begin{equation}
  H_{\rm K}
  \sin\theta
  \cos\theta
  -
  H_{x}
  \cos\theta
  \cos\varphi
  -
  H_{\rm s}
  \cos\varphi
  =
  0, 
  \label{eq:condition_2_rev}
\end{equation}
where $\cos\varphi$ in Eq. (\ref{eq:condition_2_rev}) 
should be $-1$ ($+1$) for $\varphi=\pi$ ($0$). 
Solving Eqs.(\ref{eq:condition_1_rev}) and (\ref{eq:condition_2_rev}) with $\delta\varphi \neq 0$, 
the current density necessary to keep the magnetization mostly in the stable region is given by (see Appendix \ref{sec:AppendixB})
\begin{equation}
  j_{\rm th\pm}
  =
  \frac{2eMd}{\hbar \vartheta}
  H_{x}
  \sqrt{
    \frac{2}{-h^{2} \pm h \sqrt{4+h^{2}}}
  },
  \label{eq:jth}
\end{equation}
where $j_{\rm th+(-)}$ is defined for $h>(<)0$ 
The current density $j_{\rm th}$ determines the boundary between $|m_{y}| \simeq 0$ and $|m_{y}| \gg 0$, 
which corresponds to the boundary between the regions II and III in Fig. \ref{fig:fig2}. 
In fact, Eq. (\ref{eq:jth}) well explains the boundary found by the numerical simulation, as shown by the dotted lines in Figs. \ref{fig:fig2} and \ref{fig:fig3}. 


The difference between Eqs. (\ref{eq:jc}) and (\ref{eq:jth}) is as follows. 
Equation (\ref{eq:jc}) is the threshold current density to keep the magnetization in the north hemisphere. 
When the current magnitude becomes larger than $|j_{\rm c}|$, the magnetization moves to the south hemisphere. 
However, Eq. (\ref{eq:jc}) does not provide any information as to whether $m_{y}$ becomes small or large in the steady state. 
For a switching, the magnetization in the steady state should satisfy $|m_{y}|\simeq 0$. 
Equation (\ref{eq:jth}) determines the boundary between $|m_{y}| \simeq 0$ and $|m_{y}| \gg 0$ (see Appendix \ref{sec:AppendixB}).



\begin{figure*}
\centerline{\includegraphics[width=2.0\columnwidth]{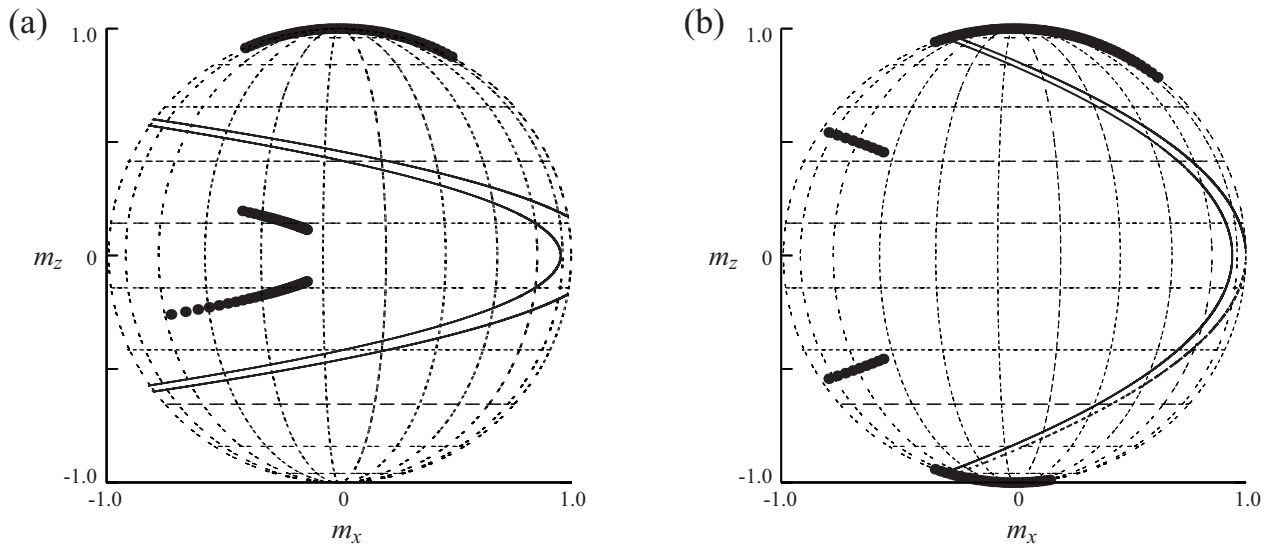}}
\caption{
         Steady state solutions in the presence of the current are shown by dots for (a) $H_{x}=50$ Oe and (b) $200$ Oe. 
         The constant energy curves near the boundary between the stable and unstable regions are shown by lines. 
         The current changes from $-200$ to $200\times 10^{6}$ A/cm${}^{2}$. 
         \vspace{-3ex}}
\label{fig:fig5}
\end{figure*}




The above results clearly indicate the theoretical conditions for the magnetization switching. 
The switching occurs when the steady-state solution in the presence of the current is in region II, as can be understood from Figs. \ref{fig:fig2} and \ref{fig:fig3}. 
The boundary between region I and other regions is determined by Eq. (\ref{eq:jc}), 
whereas the boundary between region II and III is given by Eq. (\ref{eq:jth}). 
Therefore, the switching is achieved when the current density $j$ is in the range of $j_{\rm c} \le j \le j_{\rm th}$. 
Note that the condition $j_{\rm c}<j_{\rm th}$ is satisfied when 
\begin{equation}
  |h|
  >
  0.147547...
  \simeq 0.15. 
  \label{eq:critical_field}
\end{equation}
Therefore, the magnitude of the applied field should be larger than 15\% of the perpendicular magnetic anisotropy field for the switching to take place. 
It should be emphasized that the results obtained here are applicable even to materials with low damping constant. 


The existence of the critical field, Eq. (\ref{eq:critical_field}), for the determining switching can also be understood from another viewpoint. 
The dots depicted in Fig.s \ref{fig:fig5}(a) and \ref{fig:fig5}(b) show the steady state solutions in the presence of the current on the unit sphere, 
where the longitudinal field is (a) $H_{x}=50$ Oe and (b) $200$ Oe. 
The constant energy curves near the boundary between the stable and unstable states are also shown by lines. 
Note that the critical field in the present system is $H_{\rm K}\times 0.147547...\simeq 80$ Oe. 
For $H_{x}=50$ Oe, corresponding to the field less than the critical field, 
the steady-state solutions are inside the unstable region or near the initial state ($\mathbf{m}\simeq+\mathbf{e}_{z}$). 
In such a case, a deterministic switching does not occur, as discussed in Sec. \ref{sec:Switching mechanism}. 
On the other hand, for $H_{x}=200$ Oe, which is larger than the critical value, 
some of the steady-state solutions are located outside the unstable state and near the switched state ($\mathbf{m}\simeq -\mathbf{e}_{z}$). 
The deterministic switching becomes possible in this situation.


\section{Conclusion}
\label{sec:Conclusion}

In conclusion, the comprehensive theory for achieving magnetization switching of the perpendicular ferromagnet by the spin Hall effect was developed. 
The numerical simulation of the Landau-Lifshitz-Gilbert equation indicated that 
the switching occurs when the steady state of the magnetization in the presence of the current stays in the energetically stable state. 
The theoretical formula of a threshold current was derived, which determines the boundary of the deterministic switching. 
The formula revealed that the magnitude of the magnetization field applied to the ferromagnet should be 
larger than 15\% of the perpendicular magnetic anisotropy field.


\section*{Acknowledgement}

The author is grateful to Masamitsu Hayashi and Kyung-Jin Lee for valuable discussion. 
The author is also thankful to Satoshi Iba, Aurelie Spiesser, Hiroki Maehara, and Ai Emura for their support and encouragement. 



\appendix

\section{Derivation of $j_{\rm c}$}
\label{sec:AppendixA}

In this section, the derivation of $j_{\rm c\pm}$ in the main text is shown. 
For this purpose, it is useful to express the Landau-Lifshitz-Gilbert (LLG) equation in terms of the zenith and azimuth angle, $(\theta,\varphi)$, as 
\begin{equation}
  \frac{d \theta}{dt}
  =
  -\frac{\gamma}{M \sin\theta}
  \frac{\partial E}{\partial \varphi}
  -
  \gamma
  H_{\rm s}
  \frac{\partial}{\partial \theta}
  \mathbf{m}
  \cdot
  \mathbf{p}
  -
  \alpha
  \sin\theta
  \frac{d\varphi}{dt},
  \label{eq:LLG_theta}
\end{equation}
\begin{equation}
  \sin\theta
  \frac{d\varphi}{dt}
  =
  \frac{\gamma}{M}
  \frac{\partial E}{\partial \theta}
  -
  \frac{\gamma H_{\rm s}}{\sin\theta}
  \frac{\partial}{\partial\varphi}
  \mathbf{m}
  \cdot
  \mathbf{p}
  +
  \alpha
  \frac{d\theta}{dt},
  \label{eq:LLG_varphi}
\end{equation}
where the direction of the spin polarization $\mathbf{p}$ is the unit vector in the $y$ direction in the present case. 
The energy density is given by 
\begin{equation}
  E
  =
  -MH_{x}
  m_{x}
  -
  \frac{MH_{\rm K}}{2}
  m_{z}^{2}.
  \label{eq:energy}
\end{equation}
Therefore, the steady state solutions of $(\theta,\varphi)$, satisfying $d\theta/dt=0$ and $d\varphi/dt=0$, are determined by Eqs. (\ref{eq:condition_1}) and (\ref{eq:condition_2}). 

Reference [\onlinecite{lee13}] assumes $\varphi=0$ in Eq. (\ref{eq:condition_1}). 
This assumption is valid for the dynamics before the magnetization reaches to the steady state 
with positive current $j>0$ and field $H_{x}>0$. 
However, the steady-state solution satisfies $\varphi=\pi$, as can be seen in Fig. \ref{fig:fig2}(a). 
In addition, for the negative field case $H_{x}<0$, the dynamics to the steady state satisfies $\varphi=\pi$. 
Therefore, instead of assuming $\varphi=0$, we introduce a parameter $p=\pm 1$ and reconstruct Eq. (\ref{eq:condition_2}) as 
\begin{equation}
  p
  H_{\rm K}
  \sin\theta
  \cos\theta
  -
  H_{x}
  \cos\theta
  =
  H_{\rm s}.
  \label{eq:condition_2_p}
\end{equation}
Note that a function 
\begin{equation}
  f
  =
  \left(
    p
    \sin\theta
    -
    h
  \right)
  \cos\theta,
\end{equation}
with $h=H_{x}/H_{\rm K}$ has local minima and maxima at 
\begin{equation}
  \sin\theta
  =
  \frac{1}{4}
  \left(
    ph
    +
    \sqrt{
      h^{2}
      +
      8
    }
  \right), 
\end{equation}
where we use $\sin\theta \ge 0$ because, in the spherical coordinate, $0 \le \theta \le \pi$. 
Using the solution of $\sin\theta$, we also find that 
\begin{equation}
  \cos^{2}\theta
  =
  \frac{4-h^{2}-ph \sqrt{h^{2}+8}}{8}.
\end{equation}
Substituting these solutions of $\sin\theta$ and $\cos\theta$ to the left-hand side of Eq. (\ref{eq:condition_2_p}), 
the maximum and minimum current densities $j(\propto H_{\rm s})$ satisfying Eq. (\ref{eq:condition_2}) are, depending on the values of $p=\cos\varphi=\pm 1$ and $\cos\theta$, given by 
\begin{widetext}
\begin{align}
  j_{\rm c1}
  =
  \frac{2eMd}{\hbar \vartheta}
  H_{\rm K}
  \frac{\left( -3h + \sqrt{8+h^{2}} \right)}{16}
  \sqrt{
    8
    -
    2 h 
    \left(
      h
      +
      \sqrt{
        8
        +
        h^{2}
      }
    \right)
  },
&&
  \left(
    \cos\varphi=1,\ \cos\theta>0
  \right),
  \label{eq:jc1}
\end{align}
\begin{align}
  j_{\rm c2}
  =
  \frac{2eMd}{\hbar \vartheta}
  H_{\rm K}
  \frac{\left( 3h + \sqrt{8+h^{2}} \right)}{16}
  \sqrt{
    8
    +
    2 h 
    \left(
      -h
      +
      \sqrt{
        8
        +
        h^{2}
      }
    \right)
  },
&&
  \left(
    \cos\varphi=-1,\ \cos\theta<0
  \right),
  \label{eq:jc2}
\end{align}
\begin{align}
  j_{\rm c3}
  =
  -\frac{2eMd}{\hbar \vartheta}
  H_{\rm K}
  \frac{\left( -3h + \sqrt{8+h^{2}} \right)}{16}
  \sqrt{
    8
    -
    2 h 
    \left(
      h
      +
      \sqrt{
        8
        +
        h^{2}
      }
    \right)
  },
&&
  \left(
    \cos\varphi=1,\ \cos\theta<0
  \right),
  \label{eq:jc3}
\end{align}
\begin{align}
  j_{\rm c4}
  =
  -\frac{2eMd}{\hbar \vartheta}
  H_{\rm K}
  \frac{\left( 3h + \sqrt{8+h^{2}} \right)}{16}
  \sqrt{
    8
    +
    2 h 
    \left(
      -h
      +
      \sqrt{
        8
        +
        h^{2}
      }
    \right)
  },
&&
  \left(
    \cos\varphi=-1,\ \cos\theta>0
  \right).
  \label{eq:jc4}
\end{align}
\end{widetext}
Since we study the magnetization switching from the initial state close to $+\mathbf{e}_{z}$, 
the solutions of $j_{\rm c}$ for $\cos\theta>0$ give the condition for the magnetization switching. 
In fact, Eq. (\ref{eq:jc1}) and (\ref{eq:jc4}) above are $j_{\rm c+}$ and $j_{\rm c-}$ in Eq. (\ref{eq:jc}), respectively. 

According to the derivation above, $j_{\rm c}$ determines the current density 
necessary to keep the magnetization in the north ($\cos\theta>0$) or south ($\cos\theta<0$) hemisphere with $m_{y}=0$. 
Reference [\onlinecite{lee13}] suggested that, above $j_{\rm c}$, the magnetization abruptly moves to the region of $m_{y} \neq 0$. 
We should, however, note that the current density $j_{\rm c}$ does not provide any information on the value of $m_{y}$. 
For example, let us consider the case that the magnetization initially stays in the north hemisphere. 
When the current density becomes larger than $j_{\rm c1}$, the magnetization cannot stay in the north hemisphere and moves to the south sphere. 
This is the instability threshold studied in Ref. [\onlinecite{lee13}]. 
However, the instability in the north hemisphere does not necessarily mean $m_{y} \neq 0$. 
In fact, as studied in Fig. \ref{fig:fig2}, the magnetization can stay in the south hemisphere with $m_{y}=0$. 
If the current is turned off in this situation, the magnetization relaxes to the stable state close to the south pole, as a result of the relaxation dynamics. 
Therefore, to study the accuracy of the switching, it is necessary to study whether $|m_{y}|\simeq 0$ or $|m_{y}| \gg 1$ after the magnetization moves to the south hemisphere. 
This boundary is determined by $j_{\rm th}$ found in the present work. 
The accurate switching occurs when the current density is in the range of $j_{\rm c}<j\le j_{\rm th}$. 
When the current density is larger than $j_{\rm th}$, the magnetization moves to the region of $|m_{y}| \gg 1$ because 
the spin polarization of the spin current generated by the spin Hall effect points to the $y$ direction. 
The magnetization then stays in an energetically stable state, resulting in a complex switching behavior after the current is turned off. 


\section{Derivation of $j_{\rm th}$}
\label{sec:AppendixB}

In this section, the derivation of $j_{\rm th\pm}$ in the main text is described. 
Let us consider a small perturbation, $\delta\varphi$, from $\varphi=0,\pi$ in the LLG equation. 
Equations (\ref{eq:condition_1}) and (\ref{eq:condition_2}) are rewritten as, 
\begin{equation}
  \left(
    H_{x}
    +
    H_{\rm s}
    \cos\theta
  \right)
  \delta
  \varphi
  =
  0,
  \label{eq:condition_1_rev}
\end{equation}
\begin{equation}
  H_{\rm K}
  \sin\theta
  \cos\theta
  -
  p
  H_{x}
  \cos\theta
  -
  p
  H_{\rm s}
  =
  0, 
  \label{eq:condition_2_rev}
\end{equation}
where $p=\cos\varphi\simeq \pm 1$, up to the first order of $\delta\varphi$, is introduced in Appendix \ref{sec:AppendixA}. 
Note that we are interested in the steady state solutions satisfying $\delta\varphi \neq 0$. 
Using Eq. (\ref{eq:condition_1_rev}), Eq. (\ref{eq:condition_2_rev}) can be rewritten as 
\begin{equation}
  H_{\rm K}
  \left(
    1
    -
    \sin^{2}\theta
  \right)
  +
  p H_{x}
  \sin\theta
  =
  0.
\end{equation}
Therefore, the steady state solution of $\sin\theta$ with $\delta\varphi \neq 0$ is given by 
\begin{equation}
  \sin\theta
  =
  \frac{ph + \sqrt{h^{2}+4}}{2},
\end{equation}
where we use $\sin\theta \ge 0$ again for the spherical coordinate. 
The current densities satisfying $\delta\varphi \neq 0$ and $|\delta\varphi| \ll 1$ are, depending on the values $p=\cos\varphi\simeq \pm 1$ and the sign of $\cos\theta$, 
obtained from Eq. (\ref{eq:condition_1_rev}) as 
\begin{widetext}
\begin{align}
  j_{\rm th1}
  =
  \frac{2 eMd}{\hbar \vartheta}
  H_{x}
  \sqrt{
    \frac{2}{-h^{2}+h \sqrt{4+h^{2}}}
  },
&&
  \left(
    \varphi=\pi,\ \cos\theta<0
  \right),
  \label{eq:jth1}
\end{align}
\begin{align}
  j_{\rm th2}
  =
  \frac{2 eMd}{\hbar \vartheta}
  H_{x}
  \sqrt{
    \frac{2}{-h^{2}-h \sqrt{4+h^{2}}}
  },
&&
  \left(
    \varphi=0,\ \cos\theta<0
  \right),
  \label{eq:jth2}
\end{align}
\begin{align}
  j_{\rm th3}
  =
  -\frac{2 eMd}{\hbar \vartheta}
  H_{x}
  \sqrt{
    \frac{2}{-h^{2}+h \sqrt{4+h^{2}}}
  },
&&
  \left(
    \varphi=\pi,\ \cos\theta>0
  \right),
  \label{eq:jth3}
\end{align}
\begin{align}
  j_{\rm th4}
  =
  -\frac{2 eMd}{\hbar \vartheta}
  H_{x}
  \sqrt{
    \frac{2}{-h^{2}-h \sqrt{4+h^{2}}}
  },
&&
  \left(
    \varphi=0,\ \cos\theta>0
  \right).
  \label{eq:jth4}
\end{align}
\end{widetext}
When $\cos\theta<0$, the magnetization after turning off the current relaxes to the switched state close to $-\mathbf{e}_{z}$. 
Therefore, Eqs. (\ref{eq:jth1}) and (\ref{eq:jth2}) provide the accurate switching condition. 
In fact, $j_{\rm th1}$ and $j_{\rm th2}$ defined above are $j_{\rm th+}$ and $j_{\rm th-}$ in Eq. (\ref{eq:jth}), respectively. 





\end{document}